\documentclass[10pt, conference, letterpaper]{IEEEtran}
\usepackage[utf8]{inputenc}
\usepackage{cite}
\usepackage{amsmath,amssymb,amsfonts}
\usepackage{eucal}
\usepackage{graphicx}
\usepackage{textcomp}
\usepackage{xcolor}
\def\fu#1{\textcolor{black}{#1}}

\usepackage[linesnumbered,lined, boxruled, vlined,noend]{algorithm2e}
\usepackage{algorithmic}
\usepackage{setspace}
\usepackage{booktabs}
\usepackage[hyphens]{url}
\usepackage{hyperref}
\usepackage{bm}
\usepackage{listings}
\usepackage{color}
\usepackage[inline]{enumitem}
\usepackage[normalem]{ulem}
\usepackage{multirow}
\usepackage{subcaption}
\SetKwInOut{Parameter}{parameter}

\let\oldnl\nl
\newcommand{\nonl}{\renewcommand{\nl}{\let\nl\oldnl}}
    
\begin{document}

\title{5G MEC Computation Handoff for Mobile Augmented Reality}

\author{Pengyuan Zhou*, Shuhao Fu, Benjamin Finley, Xuebing Li, \\ Sasu Tarkoma, Jussi Kangasharju, Mostafa Ammar, Pan Hui} 

\maketitle

\begin{abstract}
The combination of 5G and Multi-access Edge Computing (MEC) can significantly reduce application delay by lowering transmission delay and bringing computational capabilities closer to the end user. Therefore, 5G MEC could enable excellent user experience in applications like Mobile Augmented Reality (MAR), which are computation-intensive, and delay and jitter-sensitive. However, existing 5G handoff algorithms often do not consider the computational load of MEC servers, are too complex for real-time execution, or do not integrate easily with the standard protocol stack. Thus they can impair the performance of 5G MEC.
To address this gap, we propose \textit{Comp-HO}, a handoff algorithm that finds a local solution to the joint problem of optimizing signal strength and computational load. Additionally, \textit{Comp-HO} can easily be integrated into current LTE and 5G base stations thanks to its simplicity and standard-friendly deployability. Specifically, we evaluate \textit{Comp-HO} through a custom \textit{NS-3} simulator which we calibrate via MAR prototype measurements from a real-world 5G testbed. We simulate both \textit{Comp-HO} and several classic handoff algorithms. The results show that, even without a global optimum, the proposed algorithm still significantly reduces the number of large delays, caused by congestion at MECs, at the expense of a small increase in transmission delay.
\end{abstract}

\section{Introduction}
\label{sec:intro}
Cellular networks are a vital part of modern society with novel network technologies enabling an expanding array of use cases from simple NB-IoT sensors to immersive mobile virtual reality. Fifth-generation mobile networks (5G) specifically provide support for much higher frequencies (up to 52.6\,Ghz) with larger bandwidths (up to 400\,Mhz) and lower radio access network delay (around 10\,ms) in comparison to LTE.

Relatedly, Multi-access Edge Computing (MEC), another novel networking technology, supports deploying compute nodes near existing network nodes in the mobile network structure (often as servers co-located with base stations in the radio access network). Thus user applications that require computation can lower total delay by sending the computation request to a physically and hop-wise closer compute node rather than to a remote cloud server~\cite{5280678,10.1145/3240508.3240561}. MEC is particularly suitable for computation-intensive and delay and jitter -sensitive applications such as mobile augmented reality (MAR)~\cite{yuan2019visual,chatzopoulos2016readme,10.1145/3300061.3300116}. 

MEC has recently been standardized by ETSI thus detailing the potential for the tight integration of MEC and 5G technologies. In such a context, the handoff process~\cite{tripathi} between base stations is an important and growing concern. This is because such a handoff often means the user will also be served by a different MEC server. Under the assumption that different MEC servers have varying capabilities and loads, the user application performance will thus be affected by the capability and load of the new MEC server. However, current handoff decisions are typically based primarily on communication measurements such as signal strength, without concern for the status of MEC servers. Therefore, a handoff that improves signal strength could still reduce overall application quality due to load differences of MEC servers. This concern is especially important in 5G networks given their typically small cell sizes (\textless 500m) which implies more frequent handoffs. We denote this issue as the \textit{MEC HO} problem.

To address this issue, in this work we propose \textit{Comp-HO, a low-complexity, stand-friendly handoff algorithm jointly considering received signal quality from base stations and the computational loads of the co-located MEC servers.} In other words, when the signal strength degrades sufficiently or the serving MEC server is sufficiently overloaded, the base station initiates a handoff and re-assigns the communication and computation processes to another base station and MEC server. We focus specifically on the MAR use case, which has strict real-time quality requirements. Furthermore, we show that our approach strikes a good balance between the signal strength and computational load concerns with large numbers of MAR users.

\begin{figure}[ht]
  \centering
  \includegraphics[width=.95\linewidth]{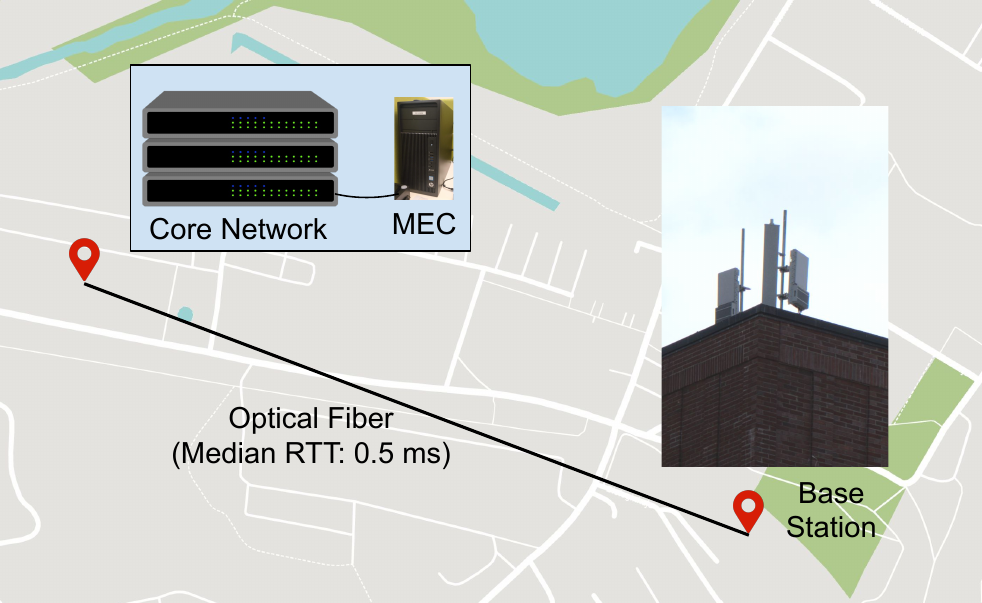}
  \caption{The 5G testbed is composed of the core network and the base station, which are interconnected via an optical fiber. The core network hosts the MEC via its NG6 interface, which is designed for connecting to the data network. The base station has two 5G antennas and one LTE antenna (middle). }
  \label{fig:5gtn_cell_tower}
\end{figure}

Specifically, we first develop a MAR prototype and deploy the prototype in a 5G MEC testbed (shown in~\autoref{fig:5gtn_cell_tower}) to measure baseline performance with a traditional handoff algorithm.\footnote{Unfortunately, the commercial 5G base station does not allow reprogramming so we cannot test the \textit{Comp-HO} algorithm in the testbed.} We then utilize the measurement results to drive a custom MEC-enabled \texttt{NS-3} simulator and compare the \textit{Comp-HO} algorithm to classic handoff algorithms on larger scale network simulations.
To the best of our knowledge, ours is one of the first efforts to address the 5G \textit{MEC HO} problem. Our contributions are threefold:
\begin{enumerate}
    \item Defining the \textit{MEC HO} problem in 5G and proposing an algorithmic solution, \textit{Comp-HO}, that jointly considers signal strength and computational load. \textit{Comp-HO} is simple and standard-friendly and outperforms traditional algorithms with only a small transmission overhead. 
    \item Measuring baseline performance with a MAR prototype in a real-world 5G MEC testbed with a traditional handoff algorithm. This baseline helps guide the parameter selection for the custom \texttt{NS-3} simulations.
    \item Carrying out reproducible and measurement-based simulations to evaluate \textit{Comp-HO} at scale using a custom MEC-enabled \texttt{NS-3} network simulator~\cite{source2020}.
\end{enumerate}

The remainder of this paper is organized as follows. Section~\ref{sec:related} presents related work in the areas of edge offloading and handoff algorithms. We describe mathematically the \textit{MEC HO} problem in Section~\ref{sec:problem} and propose \textit{Comp-HO} algorithm in Section~\ref{sec:mechanism}. Section~\ref{sec:5g} details the baseline measurements with the 5G testbed. Section~\ref{sec:simulation} presents the simulation setup and results. We discuss the potential future directions of the work in Section~\ref{sec:limit} and conclude in Section~\ref{sec:conclusion}.

\section{Related Work}
\label{sec:related}
Handoffs can be classified into horizontal and vertical~\cite{saha2004mobility,shen2006novel}, or hard and soft~\cite{carter1995evaluation,sipila1999soft}. In this work, we focus on horizontal hard handoff since we target a pure 5G networking environment and hard handoff is more common with LTE~\cite{5634584}. 

Traditional works largely make handoff decisions based on UE measurements consisting of signal quality indicators such as received signal strength~\cite{moon2009efficient,xu2013efficient}, signal-to-interference-plus-noise ratio~(SINR)~\cite{xu2010user,chowdhury2009handover} and reference signal received quality (RSRQ)~\cite{kurjenniemi,yiu2015user}. However, given the novelty of MEC, the mentioned algorithms only focus on transmission metrics without taking edge computation into consideration. The lack of such algorithms could be partly because the radio access network delay of LTE does not support some MEC-assisted applications \cite{epc} compared to novel standards like 5G, therefore lessening the motivation for LTE + MEC research.

As MEC and 5G techniques evolve, there has been some MEC-aware handoff research. Nasrin et al.~\cite{nasrin2019joint} propose a handoff algorithm that jointly considers signal quality and computational loads.
Sardellitti et al.~\cite{sardellitti2015joint} and Mao et al.~\cite{mao2017joint} focus on the joint optimization of radio and computational resources considering energy consumption and user experienced delay.\footnote{In this work, we define user experienced delay as the time between when user sends a request and receives a reply.} Basic et al.~\cite{basic2019fuzzy} propose a fuzzy logic handoff algorithm that selects a target node based on bandwidth, processor, and delay parameters of edge servers. Emara et al.~\cite{emara2018mec} and Li et al.~\cite{li2017efficient} both propose to improve handoff algorithms by considering MEC load in 5G heterogeneous networks and cloud radio access networks, respectively. 

Ma et al.~\cite{ma2017efficient} propose to build an efficient service handoff system across edge servers based on Docker container migration. Wang et al.~\cite{wang2018user} utilize a lightweight heuristic algorithm to reduce offloading task execution delay by jointly considering task information, small base station and user mobility information. Yu et al.~\cite{yu2018dmpo} propose a dynamic algorithm for partial offloading based on short-term mobility prediction to minimize energy consumption while satisfying delay requirements.

To summarize, we find the existing MEC-aware handoff works \textbf{fall short in several respects}:
\begin{enumerate}
    \item Most solutions do not provide realistic algorithms that take the X2 application protocol into consideration~\cite{nasrin2019joint,basic2019fuzzy,emara2018mec,li2017efficient,ma2017efficient,wang2018user,yu2018dmpo}.
    \item Some of the proposed algorithms have computational complexities larger than $O(n^2)$~\cite{wang2018user,yu2018dmpo}. In cases with larger numbers of UEs, this complexity can be problematic.
    \item Related proposals tend to require additional message transmissions between UE and base station to collect information for the handoff algorithm. This overhead impacts system performance through additional link transmissions and information collection delay.
    \item Related works do not inform their simulations with empirical 5G measurements thus making their interpretations less reliable~\cite{nasrin2019joint,basic2019fuzzy,emara2018mec,li2017efficient,ma2017efficient,wang2018user,yu2018dmpo}.
    \item Most related works lack detailed simulations. The related works conduct only numerical modeling or simplified simulations without millisecond granularity and packet-level/multi-layer detail (e.g., from physical to application layer), thus excluding some detailed dynamics only visible with such simulations~\cite{nasrin2019joint,basic2019fuzzy,emara2018mec,li2017efficient,wang2018user,yu2018dmpo}.
\end{enumerate}

\textit{We address these shortfalls by proposing an easily-deployable handoff algorithm with minimum overhead. We code the key metric values collected from a real 5G testbed together with \textit{Comp-HO} algorithm into a custom MEC-enabled \texttt{NS-3} simulator following base station protocol standard and perform a packet-level simulation.
}

\section{Problem Formulation}
\label{sec:problem}
This section formulates the \textit{MEC HO} problem in the 5G context with MEC servers co-located with 5G base stations. The formulation focuses on optimizing the performance of MEC-assisted UE applications with respect to experienced delay. Also we assume the transmission delay between a MEC server and its co-located base station is negligible (as also shown in the 5G testbed measurements, see ~\autoref{fig:5gtn_cell_tower}).

We let $\CMcal{M} \triangleq \{1,2,...,M\}$ denote the $M$ MEC servers and $\CMcal{U}=\{1,2,...,U\}$ the U mobile UEs in the system. Each MEC server has a fixed capacity $c$~(maximum queue length) and is connected directly to a 5G base station via fixed-line Ethernet. 
The time horizon is discretized into slots of equal periods indexed by $t\in \mathbb{N}$.
We note several important MEC server assumptions:
\begin{enumerate}
\item \textit{Homogeneity of MEC servers}: The MEC servers have the same data processing rates. 
\item \textit{Job-level migration}: The tasks of a job may execute on any given MEC server.
\item \textit{Task atomicity}: A task cannot be split across MEC servers.
\item \textit{Non-preemptive task scheduling}: A task being processed cannot be interrupted by any other task.
\end{enumerate}

Let $D_u^m$ denote the transmission delay from UE $u$ to MEC $m$ excluding the processing delay. In other words, $D_u^m$ consists of the uplink and downlink delay. Normally, either the uplink or downlink has larger data packet sizes and thus dominates the transmission delay. For example, uplink delay dominates the transmission delay, since the uplink packets to MEC servers contain much more data than the downlink packets for most MAR offloading applications. 
%
To simplify the problem, we consider only the dominant direction of data transmission, $D_u^m $, which solely depends on the signal quality received by $u$ from $m$, i.e., $S_u^m $:
\begin{equation}
 D_u^m =  \phi(S_u^m).
\end{equation}
where the function $\phi: R^1\rightarrow  R^1$ is monotonically decreasing.

\renewcommand{\arraystretch}{1.1}
\begin{table}[!t]
\caption{Notation Table}
\centering
\begin{tabular}{ll}
\specialrule{1.3pt}{1pt}{1pt}
$u$ & UE $u$ \\
$m$ & MEC $m$ \\
$\CMcal{S}$ & Serving base station and MEC server \\
$\CMcal{T}$ & Targeting base station and MEC server \\
$Q^m$ & Processing queue length in MEC $m$ \\
$D_u^m$ & Transmission delay from $u \text{ to } m$ \\
$S_u^m$ & Signal quality for $u \text{ received from the cell with } m$ \\
$R_u$ & UE measurement: $\langle  \mathbf{S_\mathnormal{u}^m}, A \rangle$\\
$\theta$ & RSRQ threshold\\
$\delta$ & Handoff offset\\
\specialrule{1.3pt}{1pt}{1pt}
\end{tabular}
\label{tab:notation}
\end{table}

Let $Q^m$ denote the processing queue length of MEC $m$ at a point in time. We assume the change in user experienced delay during the handoff from MEC $\CMcal{S}$ to $\CMcal{T}$ depends on the difference in signal qualities and queue lengths as follows:
\begin{equation}
\bigtriangleup D_u^{\CMcal{S},\CMcal{T}}=f(\bigtriangleup S_u^{\CMcal{S},\CMcal{T}}, \bigtriangleup Q^{\CMcal{S},\CMcal{T}}),
\label{eq:delay}
\end{equation}
where $\bigtriangleup X_u^{\CMcal{S},\CMcal{T}} = X_u^{\CMcal{T}}-X_u^{\CMcal{S}}$.
The function $f$ denotes that both signal strength and computational load are considered.

Let $a_u^{\CMcal{S},\CMcal{T}}$ indicate whether MEC $\CMcal{S}$ hands off $u$ to $\CMcal{T}$ as follows:
\begin{equation}
 a_u^{\CMcal{S},\CMcal{T}} =
\begin{cases}
  1 & \text{if $\CMcal{S}$ hands off task of $u$ to $\CMcal{T}$} , \\
  0 & \text{otherwise, including no handoff requests}.
\end{cases} \\
\end{equation}

We can then formulate an optimization problem as follows,
\begin{equation}
\begin{aligned}
\min \; \;   &\displaystyle \sum_{u \in \CMcal{U}} \bigtriangleup D_u^{\CMcal{S},\CMcal{T}} a_u^{\CMcal{S},\CMcal{T}} \\
\text{s.t.} \; \; & \CMcal{S}, \CMcal{T}\in \CMcal{M}, \CMcal{S} \neq \CMcal{T}\\
\end{aligned}	
\label{eq:problem_p1}
\end{equation}
The problem aims at optimizing user experience by minimizing overall user experienced delay for all UEs. The problem can be seen as a linear sum assignment problem which is also known as a minimum weight matching in bipartite graphs. Balanced assignment algorithms such as the Hungarian algorithm~\cite{bruff2005assignment} can be used to solve this problem. Therefore, an optimal solution to the problem with global information is possible. 

\IncMargin{1em}
\begin{algorithm}[t]
\setstretch{1.2}
\SetAlgoLined\SetArgSty{}
	\SetAlgoLined\DontPrintSemicolon
	\SetKwProg{proc}{thread}{:}{}
	\SetKwProg{main}{Main}{:}{}
	\Parameter{$S_u^m \gets \text{RSRQ $u$ received from $m$}$\\
	$Q^m \gets \text{Max queuing time in $m$}$}
	\BlankLine
	\nonl {\textbf{UE Measurement}}\\
	\proc{ReportUeMeasurment($R_u$)}{
	  \While{$u$ offloading to $m$}{
	    \textit{$\mathbf{S_\mathnormal{u}^m} \gets S_u^m$ for all probeable MECs, $\textbf{m} \subseteq \CMcal{M}$} \\
	    \textit{$R_u \gets \langle \mathbf{S_\mathnormal{u}^m}, A \rangle $} 
	    \tcp*{$A-> \textit{App info}$}
	    \textit{sendUeMeasurement}$(R_u)$\label{alg:lineue} 
	  }
	}
	\SetKwProg{proc}{thread}{:}{}
	\SetKwProg{main}{Main}{:}{}
	\BlankLine
	\BlankLine
	\nonl {\textbf{Handoff}}\\
	\proc{updateUeMeasurement($\mathbf{R_u}$)}{
	  \textit{$\mathbf{R_u} \gets R_u$ from all connected UEs, $\textbf{u}\subseteq \CMcal{U}$}\label{alg:line4} \\
	} 
	\BlankLine
	\proc{updateLoad($\mathbf{Q^m}$)}{
	  \textit{$\mathbf{Q^m}, \mathbf{A} \gets \langle Q^m,A \rangle$ from all nearby MECs, $\textbf{m}\subseteq \CMcal{M}$} \label{alg:line6} \\
	} 
    \BlankLine
	\proc{Hand off u}{
	  \If{$S_u^\CMcal{S} < \theta $}{\tcp*{$\theta$-> \textit{RSRQ threshold}\label{alg:line8}}  
	  \textit{$F(S_u^\CMcal{T},Q^\CMcal{T})=$\textit{max\_element}($F(\mathbf{S_\mathnormal{u}^m} ,\mathbf{Q^m}))$} \label{alg:line9}\\
	  \If{\textit{$F(S_u^\CMcal{T},Q^\CMcal{T}) - F(S_u^\CMcal{S},Q^\CMcal{S}) > \delta$}}{\label{alg:line10}
	  \tcp*{$\delta$-> \textit{Handoff offset} }
	  \textit{SendHoRequest}($\CMcal{T}$)\label{alg:line11}\\
	      }
	  }
	}
\caption{\textit{Comp-HO} algorithm}
\label{alg:cho}
\end{algorithm}
However, the delay overhead caused by global information collection and decision dissemination can easily worsen the system performance. Moreover, most current standards require base stations to make their own handoff decisions and numerous standardized policies would require modifications. Therefore, we instead develop \textit{Comp-HO} as a local optimization algorithm and leave global optimization for future work.
\begin{figure*}[!t]
\centerline{\includegraphics[width=0.8\textwidth]{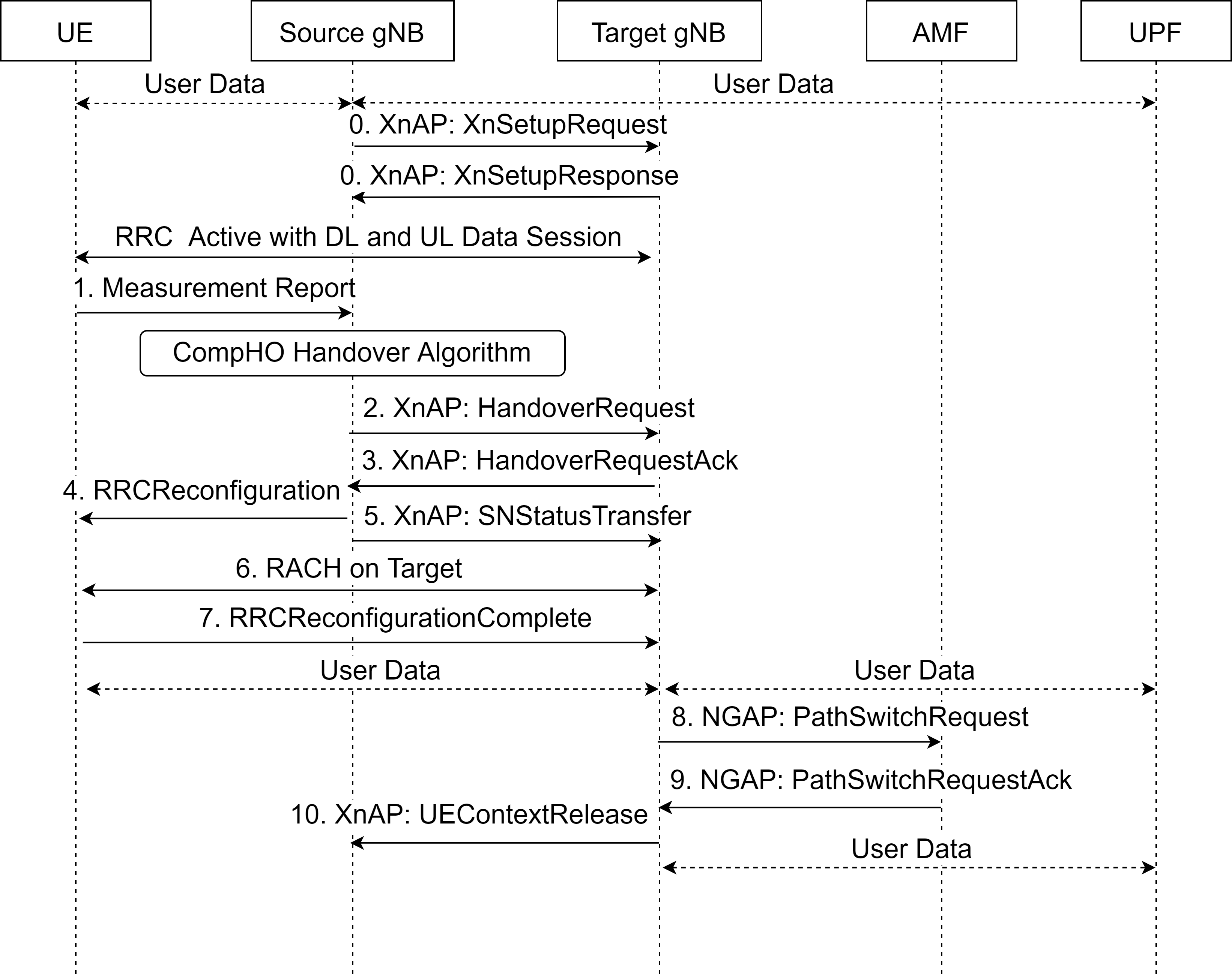}}
\caption{High-level \textit{Comp-HO} flow diagram using 3GPP 5G and ETSI MEC \cite{etsi2019} terminology. \textit{Comp-HO} specific flows are marked in blue.}
\label{fig:HOflow}
\end{figure*}
\section{Comp-HO Algorithm}
\label{sec:mechanism}
\subsection{Algorithm}
\label{subsec:alg}
To perform a \textit{MEC HO}, each UE sends measurement reports to the serving base station. The serving base station runs the handoff algorithm and decides when to initialize a handoff to a target base station. We first describe the UE measurement and then the handoff algorithm. Table~\ref{tab:notation} summarizes the related notations. Each UE collects the signal qualities~(RSRQ) from all nearby base stations and sends them to the serving base station together with the information of its MEC-assisted applications in the report as shown on line~\ref{alg:lineue} in Algorithm~\ref{alg:cho}.

In parallel, each base station collects UE measurements from connected UEs and load information from nearby MEC servers, respectively (\autoref{alg:line4} to~\ref{alg:line6}). The load information includes metadata of MEC-assisted applications running in each MEC server and the processing queue lengths. $F(S_u^m, Q^m)$ denotes the weighted sum of signal quality and queue length to perform the optimization. To keep the complexity low, $F()$ follows a linear form: $F(S_u^m, Q^m)=w_s *S_u^m- w_q *Q^m$. In the simulation, we iteratively tried different sets of weights and offsets to optimize performance. The base station starts the handoff of a UE if its RSRQ fails to meet the threshold~(\autoref{alg:line8}). Utilizing the collected load information and signal quality values, the base station selects the target base station and MEC server and sends the handoff request~(\autoref{alg:line9}-\ref{alg:line11}). The offset metric, $\delta$, is introduced to avoid the ping-pong effect~(\autoref{alg:line10}). 

\subsection{Flow Diagram}
\label{subsec:flow}
In terms of interaction between 5G network elements, Figure \ref{fig:HOflow} illustrates the high-level message flows between such elements just before and during a \textit{Comp-HO}. The figure uses the terminology and interfaces following standard 3GPP 5G and ESTI MEC~\cite{etsi2019}, except that the Network Exposure Function (NEF) or Radio Network Interface (RNI) should allow passing load information back to the gNB/ng-eNB to be used in the \textit{Comp-HO} algorithm. These NEF/RNI interfaces are currently designed to provide radio network information such as cell change notifications to the MEC system for user context (e.g., virtual machine) migration between MEC servers. 

Also, we note that the analogous flows with a 5G radio access network and LTE EPC (like in testbed network) would be only slightly different in functions and terminologies. Overall, the flows illustrate how \textit{Comp-HO} could potentially function after integrated into current mobile standards.

\subsection{Scalability}
\label{sec:scale}
The local algorithm is $O(n)$ in time complexity where $n$ is the number of sector UEs. This complexity is the same as the baselines~(single signal indicator based handoff algorithms), so the algorithm scales in time at least as well as those. 
Whereas in terms of measurement messages, in our scenario the UE does not deal with MEC load info messages directly but these messages are transferred between the MECs and base stations over Xn links (as Figure \ref{fig:HOflow} illustrates). Thus the messaging complexity scales with the number of MECs and base stations rather than UEs, thus allowing good scaling as the number of UEs increase.

\section{5G MEC Measurements}
\label{sec:5g}
\subsection{5G MEC Testbed and MAR Prototype}
\label{subsec:5gtn}
Our 5G MEC testbed is a 5G micro-operator~\cite{matinmikko2017micro} network built by a joint national effort of academic and industrial partners. As shown in \autoref{fig:5gtn_cell_tower}, the testbed is composed of two parts: a 5G Core~\cite{3gpp.21.915} and a base station. The core network is deployed in non-standalone (NSA) mode and the network functions (NFs) are implemented in Linux virtual machines located on servers in a single university server room. The MEC server is located in the same server room and connected via Ethernet to the core network switches. The base station is located on the roof of another building. The base station has two antennas for 5G and one for LTE, providing coverage over the campus area. The base station is connected to the core network via optical fiber. According to our measurements, the median RTT between the base station and MEC is 0.5 ms. 

We develop a MAR prototype by running a custom Android client app on a Huawei Mate 30 Pro 5G smartphone (the UE) and a Linux server app on the MEC equipped with 8-core Xeon CPU, 16 GB memory and Quadro K2200 GPU. The client app captures camera frames at 10 frame rate~(FPS). Then, it downscales the frames to $480\times320$ pixels and sends to the MEC. 

The MEC receives the frames and uses YOLO~\cite{redmon2016you} to perform object detection. The object detection result is composed of a set of bounding boxes of the detected objects as well as the object classes and detection confidences~\cite{yolov3}. Once the objects are detected, the result is sent back to the UE and rendered on the screen, annotating the objects from camera view.

Unfortunately, due to technical and licensing limitations, we cannot implement and deploy the \textit{Comp-HO} algorithm into the base station. Instead, we conduct network measurements to observe the baseline performance of the MAR prototype system in Section \ref{sec:scenario}. Then, in Section \ref{sec:simulation}, we use these measurement results to inform the simulation parameters thus helping to mitigate the gap between a real-world 5G network and the \texttt{NS-3} simulator.

\subsection{Measurement Setup and Results}
\label{sec:scenario}
We record the transmission delays of the frames and detection results by timestamping during the sending and receiving on the UE and MEC. We connect the UE and the MEC via Ethernet beforehand to estimate the clock drift (within a confidence interval of \textless1\,ms). We also select an off-peak time to conduct the measurements. The results, therefore, illustrate MAR performance with 5G MEC without non-MAR loads from other UEs.

\begin{figure}[!t]
  \centering
  \includegraphics[width=\linewidth]{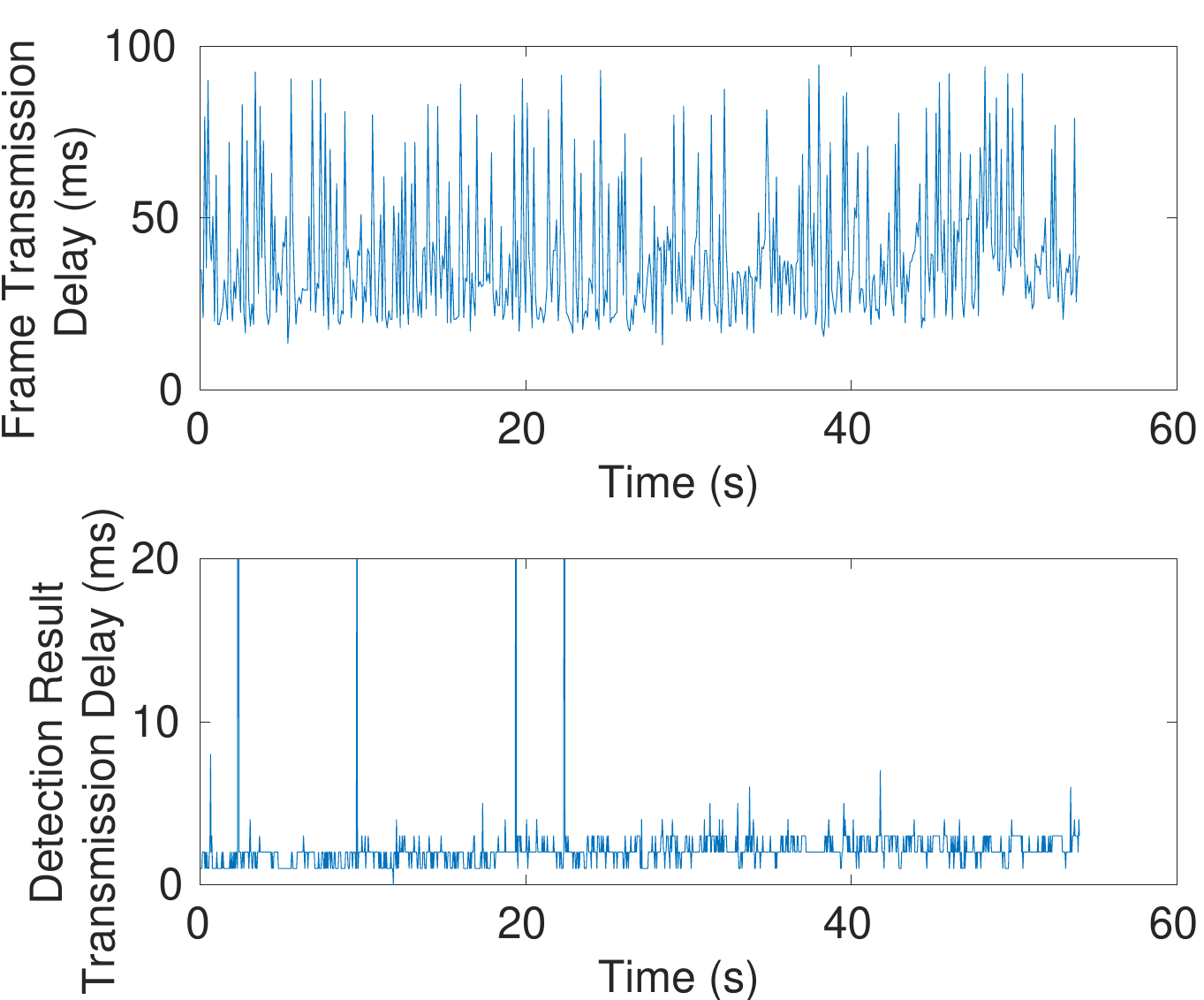}
  \caption{Frame transmission delay and detection result transmission delay between the UE and the MEC, measured with the MAR prototype in 5G MEC testbed.}
  \label{fig:server_client_delay}
\end{figure}
\renewcommand{\arraystretch}{1.2}
\begin{table}[!t]
\caption{Measurement statistics of~\autoref{fig:server_client_delay}. UE $\rightarrow$ MEC: frame transmission (uplink). MEC $\rightarrow$ UE: detection result transmission (downlink). }
\label{table:5g_testing}
\centering
\begin{tabular}{lcc}
\specialrule{1.3pt}{1pt}{1pt}
& UE $\rightarrow$ MEC & MEC $\rightarrow$ UE \\
\hline
Median delay (ms) & 32.00 & 2.00 \\ 
Jitter (ms) & 9.5 & 0 \\
Packet loss (\%) & $0.06$ & $0.52$ \\
\specialrule{1.3pt}{1pt}{1pt}
\end{tabular}
\end{table}

We walk along a fixed route during the measurement and collect the results shown in \autoref{fig:server_client_delay} and \autoref{table:5g_testing}. Overall, the UE sends 540 frames to the MEC over a period of 54 seconds. The median frame (uplink) transmission delay and the result (downlink) transmission delay are 32.0 ms and 2.0 ms, respectively. This large difference is due to two primary reasons. 
\begin{enumerate}
    \item Firstly, the uplink bandwidth is much smaller than the downlink. According to our measurements, the downlink throughput is about 360 Mbps while the uplink only 30 Mbps.
    \item Secondly, a frame has much larger size than its detection result. Specifically, a frames after compression is typically a few kilobytes while the detection result is simply plain text only typically only tens of bytes.
\end{enumerate}
The combination of these two factors contributes to a higher uplink transmission delay than downlink transmission delay.

\section{Simulation}
\label{sec:simulation}
\subsection{NS-3 Simulator and Setup}
We next perform simulations to estimate the performance of the \textit{Comp-HO} algorithm at scale. The simulation setup aligns with some important metrics and results taken from the 5G measurements such as the frequency, UE speed and lower bound of user experienced delay.

\begin{figure}[!t]
\centerline{\includegraphics[width=.8\linewidth]{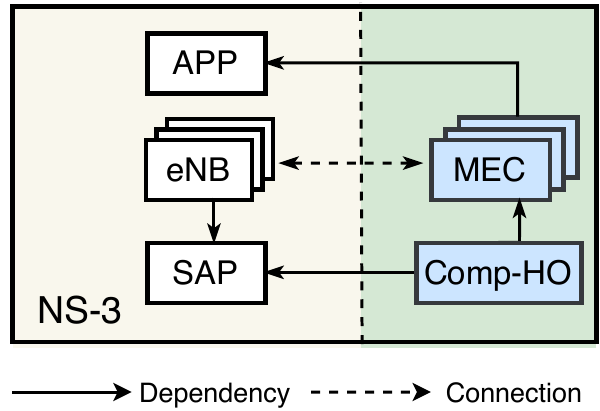}}
\caption{Custom \texttt{NS-3} simulator}
\label{fig:simulator}
\end{figure}

\vskip 0.03in\noindent\textbf{Simulator: }We modify the existing LTE module\footnote{We use the LTE module rather than a 5G \texttt{NS-3} module because handoff support in such modules \cite{patriciello2019,mezzavilla2018} is still limited.} of \texttt{NS-3} so that each eNB is co-located with three MEC servers (one for each sector) and any UE data packets are forwarded to a MEC server rather than to a packet gateway~(\autoref{fig:simulator}). We also integrate the \textit{Comp-HO} algorithm into the LTE module. Each MEC server contains several queues that each process packets at a fixed rate. For a given MEC server, the processing queue length reported to the eNB (and thus to the handoff algorithm) is the minimum length out of these queues.
\vskip 0.03in\noindent\textbf{Transmission Setup:} For simplicity, each UE sends UDP packets at a fixed FPS (20\,Hz) but with random starting times to avoid synchronization. As mentioned, the packets are forwarded by the eNB to the corresponding MEC for processing and after processing sent back to the UE. If after processing, the UE is no longer being served by the eNB that received the packet originally (for example because of handoff) the packet is discarded. We denote this type of packet loss as MEC-mobility packet loss. For performance monitoring, the round-trip packet delay (which we refer to as the UE experienced delay) for each packet is recorded at the UE.

\vskip 0.03in\noindent\textbf{Variations:} We perform both basic simulations (with the noted parameters) and several simulation variations as follows.
\begin{enumerate}
    \item \textit{Handoff rate:} We vary the UE speeds to alter the effective handoff rate. 
    \item \textit{FPS:} We vary the FPS to uniformly alter the MEC loads (given the fixed processing rate), 
    \item \textit{Mobility:} We use two different UE mobility models (one with a center bias and one without) to illustrate the impact of different spatial UE distributions (e.g., a central crowd of UEs).
\end{enumerate}

For further clarification on the mobility models, we use random waypoint as the baseline model as random waypoint has a center bias \cite{bettstetter2004} thus allowing a higher central user density and heterogeneous load across MECs. Such heterogeneity naturally helps illustrate the benefits of the \textit{Comp-HO} algorithm. However, for reference, we also use a Gauss-Markov mobility model which does not have such a center bias \cite{broyles2010}, and thus has a more homogeneous user distribution and load across MECs. We also perform each simulation three times (with different random seeds) to ensure performance differences are not due to random variation.
\begin{figure}[!t]
\centerline{\includegraphics[width=.9\linewidth]{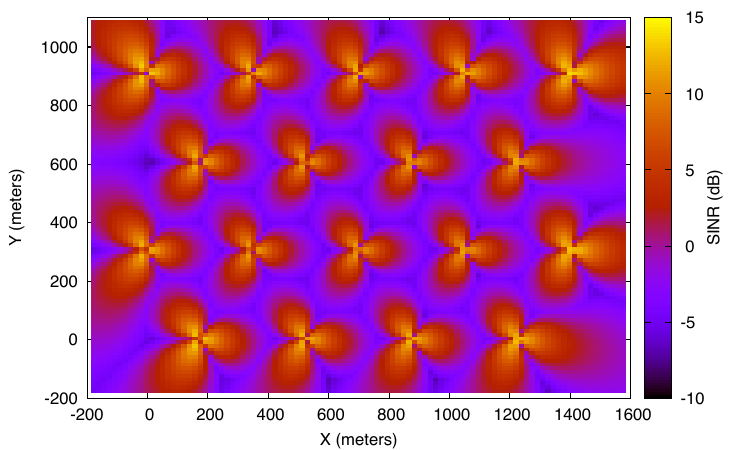}}
\caption{\texttt{NS-3} simulation network layout map with SINR}
\label{fig:sim_map}
\end{figure}

\begin{table}[!t]
\caption{\texttt{NS-3} simulation parameters}
\label{tab:sim_parameters}
\centering
\begin{tabular}{ll}
\specialrule{1.3pt}{1pt}{1pt}
Parameter & Value \\
\hline
Number of base stations & 18 Tri-sector eNBs\\
Layout & Hexagonal \\
Intersite Distance & 350 m \\
Center Frequency & 3.55 Ghz (band 22) \\
Bandwidth & 20 Mhz \\
Path Loss Model & ITU-R P.1411 LoS \cite{itu1411} \\
Height & eNB: 45 m, UE: 1.5 m \\
Number UEs & 50 \\
UE Mobility & Random Waypoint \\
UE Velocity & 2 m/s (7.2 km/h) \\
Queues per MEC Server & 1\\
Simulation Area & 1800x1300 m \\
Simulation Time & 30 s\\
\specialrule{1.3pt}{1pt}{1pt}
\end{tabular}
\end{table}
\begin{table}[!t]
\caption{Benchmark Parameters}
\label{tab:benchmark}
\centering
\begin{tabular}{ll}
\specialrule{1.3pt}{1pt}{1pt}
Algorithm & Parameters  \\ \hline
\textit{A2-A4-RSRQ} & ServingRsrqThreshold=30~\cite{3gpp.36.331} \\
                    & NeighbourRsrqOffset=1~\cite{3gpp.36.331}\\
\textit{A3-RSRP}    & TimeToTrigger=256 (ms)~\cite{3gpp.36.331} \\
                    & Hysteresis=3 (dB)~\cite{3gpp.36.331}\\
\specialrule{1.3pt}{1pt}{1pt}
\end{tabular}
\end{table}

\begin{figure*}
  \begin{subfigure}{0.24\textwidth}
    \includegraphics[width=\linewidth]{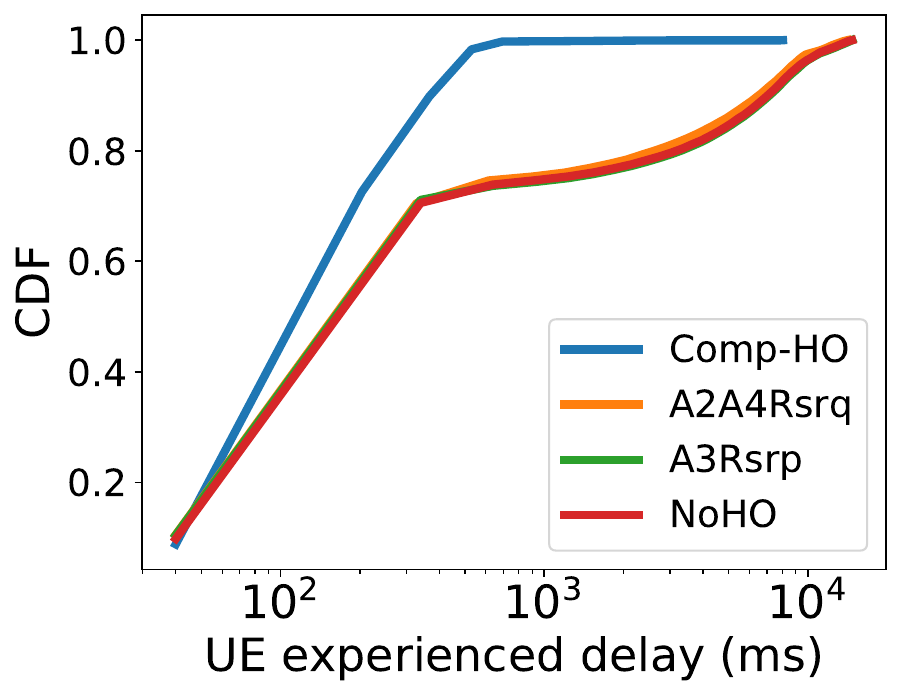}
      \caption{UE experienced delay}
      \label{fig:delay}
    \end{subfigure}
  \begin{subfigure}{0.24\textwidth}
    \includegraphics[width=\linewidth]{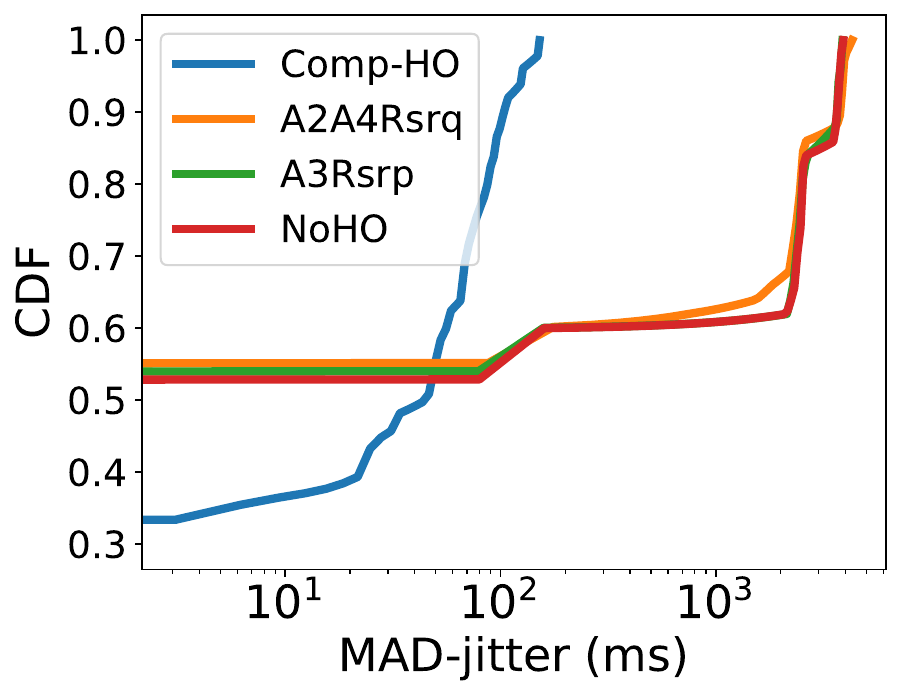}
      \caption{UE experienced MAD-jitter}
      \label{fig:jitter}
    \end{subfigure}
  \begin{subfigure}{0.24\textwidth}
    \includegraphics[width=\linewidth]{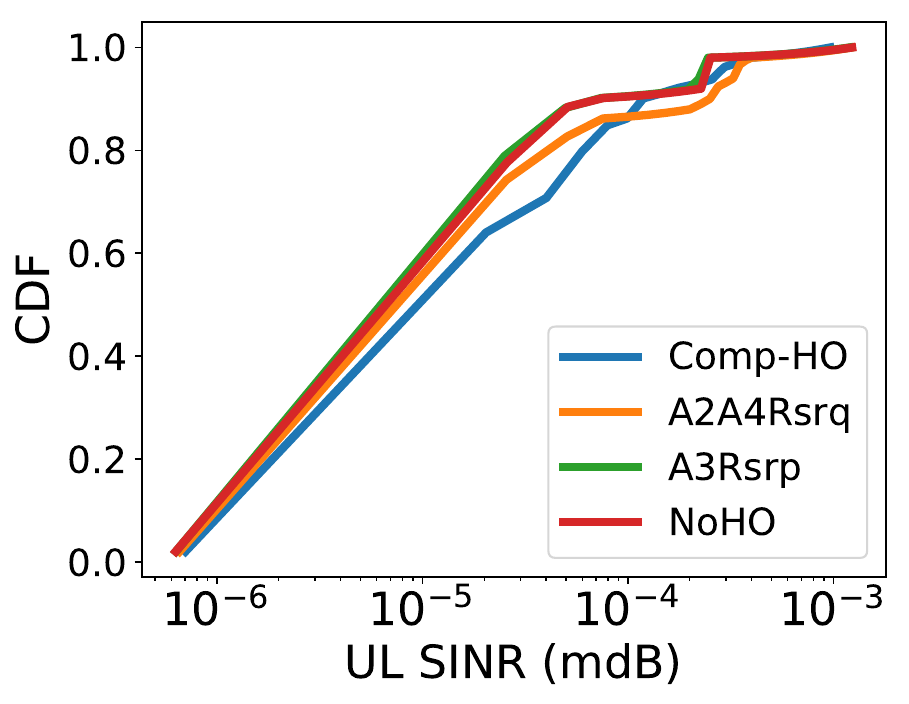}
      \caption{Uplink SINR}
      \label{fig:ulsinr}
    \end{subfigure}
  \begin{subfigure}{0.24\textwidth}
    \includegraphics[width=\linewidth]{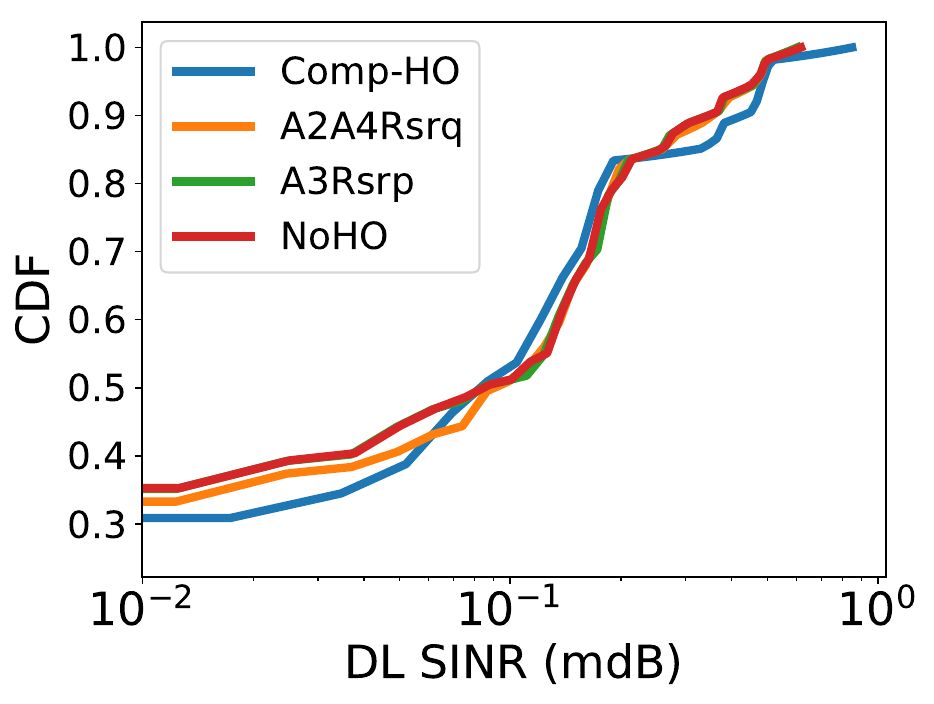}
      \caption{Downlink SINR}
      \label{fig:dlsinr}
    \end{subfigure}
    \caption{UE experienced delay, MAD-jitter, and uplink and downlink SINR (all x-axes on log scale) for \textit{Comp-HO}, \textit{A2-A4-RSRQ}, \textit{A3-RSRP}, and \textit{NoHO} (no handoff)}
    \label{fig:7}
\end{figure*}

\vskip 0.03in \noindent\textbf{Benchmark: }We compare the \textit{Comp-HO} algorithm to two existing LTE handoff algorithms (\textit{A2-A4-RSRQ} and \textit{A3-RSRP}) and to a scenario with no handoffs~(NoHO). The \textit{A2-A4-RSRQ} and \textit{A3-RSRP} algorithms are based on the A2, A4 and A3 control events defined by 3GPP standard~\cite{access2013radio}. The \textit{A2-A4-RSRQ} algorithm triggers a handoff when the serving cell's RSRQ drops below a threshold~(A2) and a neighboring cell's RSRQ rises above an offset~(A4). The \textit{A3-RSRP} algorithm triggers a handoff when the serving cell's Reference Signal Received Power~(RSRP) drops below the RSRP of a neighboring cell. Both algorithms are already part of the LTE module of \texttt{NS-3}~\cite{ns3dochandover}. The other simulation parameters are summarised in Table~ \ref{tab:sim_parameters}. The developed \texttt{NS-3} code is available at \cite{source2020}.

\begin{figure*}
  \begin{subfigure}{0.24\textwidth}
    \includegraphics[width=\linewidth]{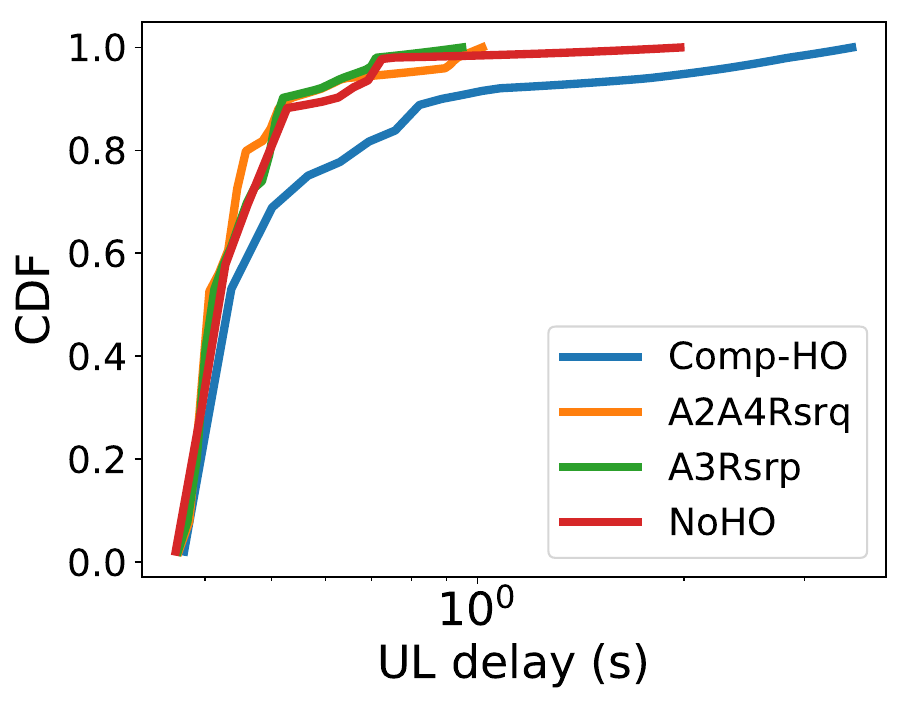}
      \caption{Uplink trans. delay}
      \label{fig:uldelay}
    \end{subfigure}
  \begin{subfigure}{0.24\textwidth}
    \includegraphics[width=\linewidth]{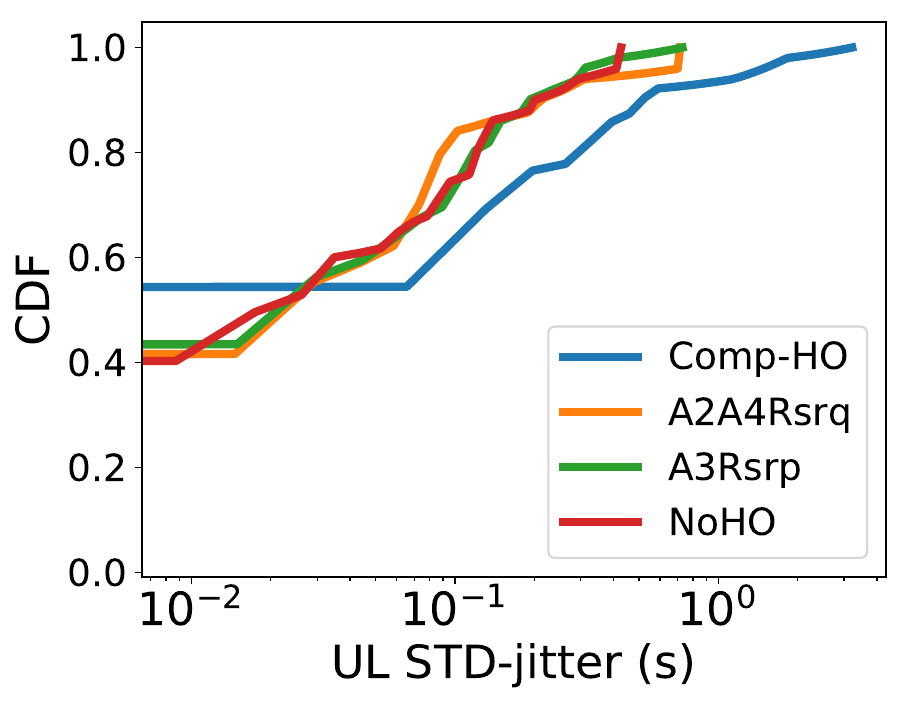}
      \caption{Uplink trans. STD-jitter}
      \label{fig:uldelayjitter}
    \end{subfigure}
  \begin{subfigure}{0.24\textwidth}
    \includegraphics[width=\linewidth]{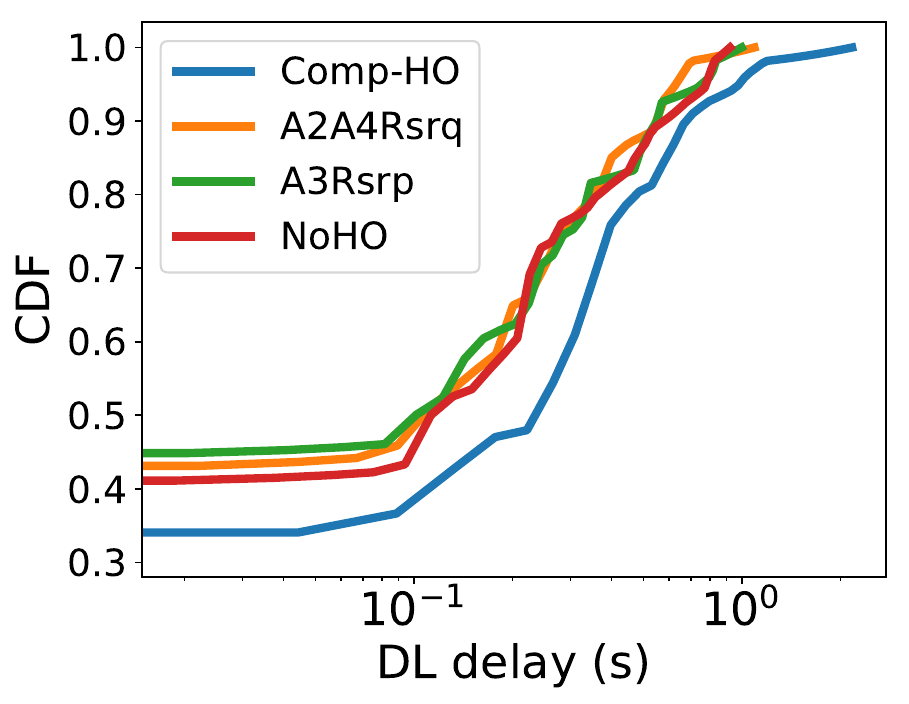}
      \caption{Downlink trans. delay}
      \label{fig:dldelay}
    \end{subfigure}
  \begin{subfigure}{0.24\textwidth}
    \includegraphics[width=\linewidth]{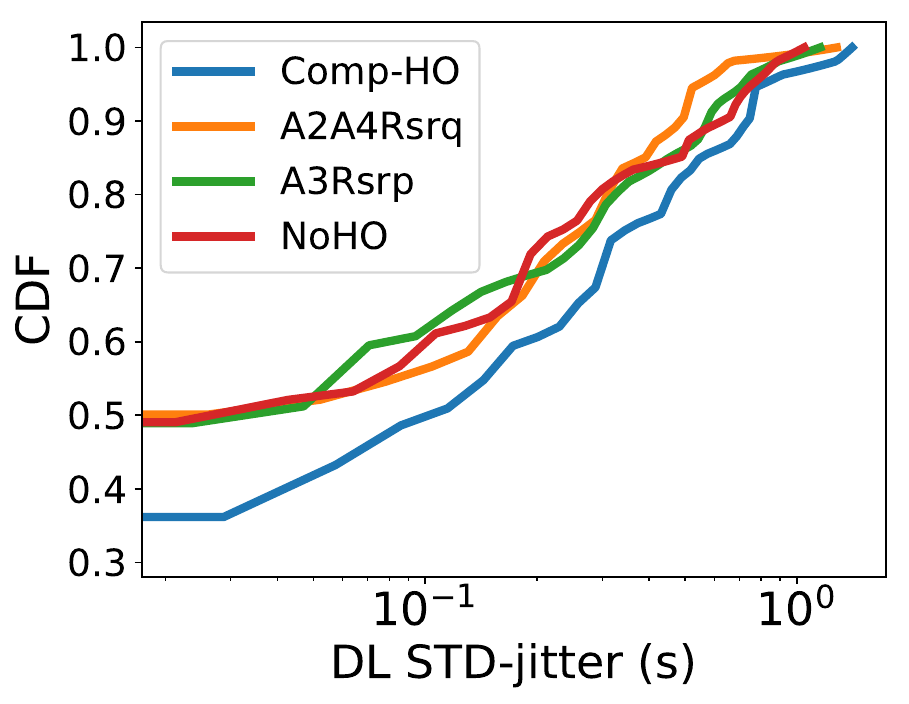}
      \caption{Downlink trans. STD-jitter}
      \label{fig:dldelayjitter}
    \end{subfigure}
    \caption{Transmission delay and STD-jitter (excludes all processing delays, all x-axes on log scale) for \textit{Comp-HO}, \textit{A2-A4-RSRQ}, \textit{A3-RSRP}, and \textit{NoHO} (no handoff)}
    \label{fig:8}
\end{figure*}
\begin{figure*}
\begin{subfigure}{0.24\textwidth}
    \includegraphics[width=\linewidth]{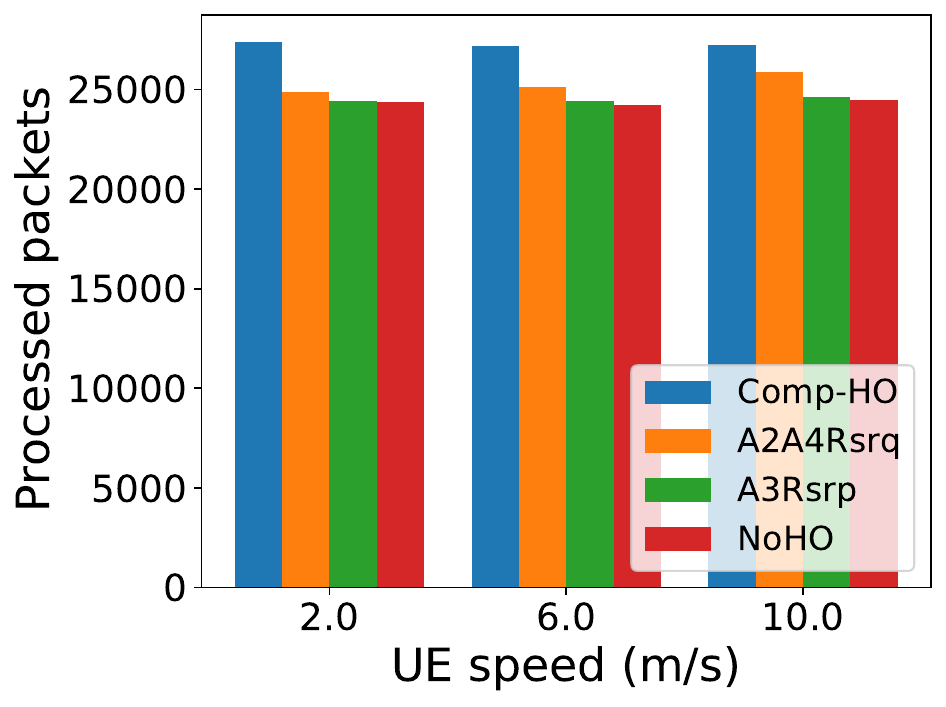}
      \caption{MEC processed packets}
      \label{fig:process}
    \end{subfigure}
  \begin{subfigure}{0.24\textwidth}
    \includegraphics[width=\linewidth]{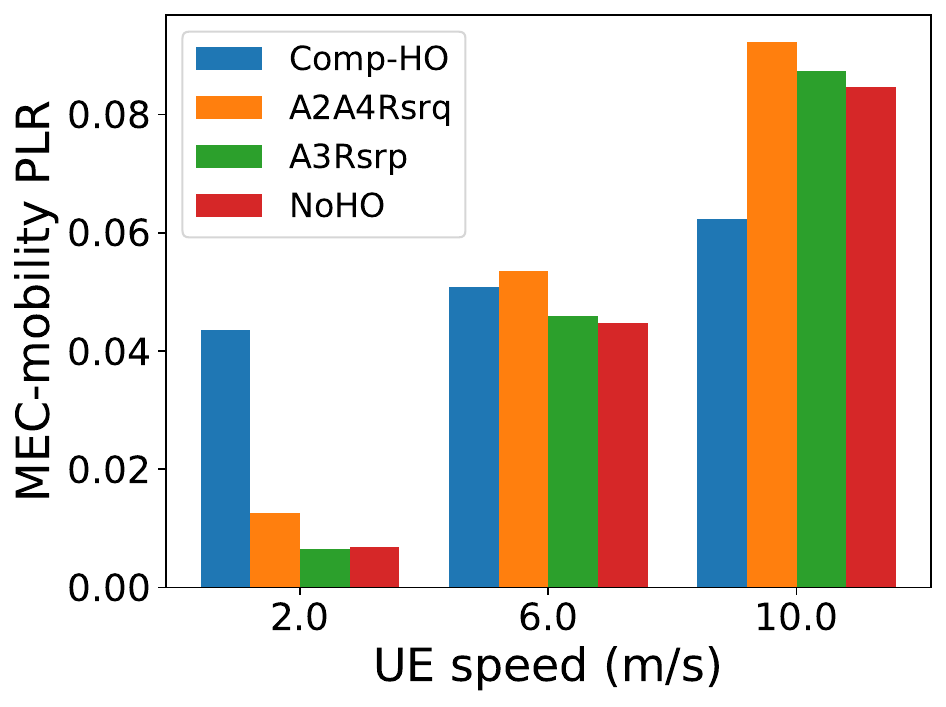}
      \caption{MEC-mobility PLR}
      \label{fig:plr}
    \end{subfigure}
  \begin{subfigure}{0.24\textwidth}
    \includegraphics[width=\linewidth]{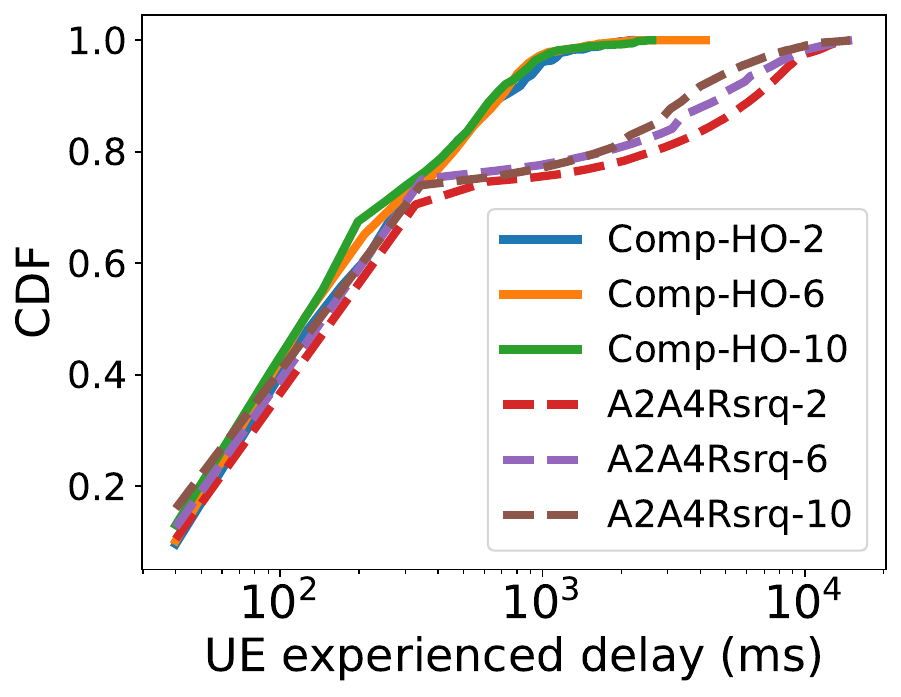}
      \caption{UE experienced delay}
      \label{fig:delayspeed}
    \end{subfigure}
  \begin{subfigure}{0.24\textwidth}
    \includegraphics[width=\linewidth]{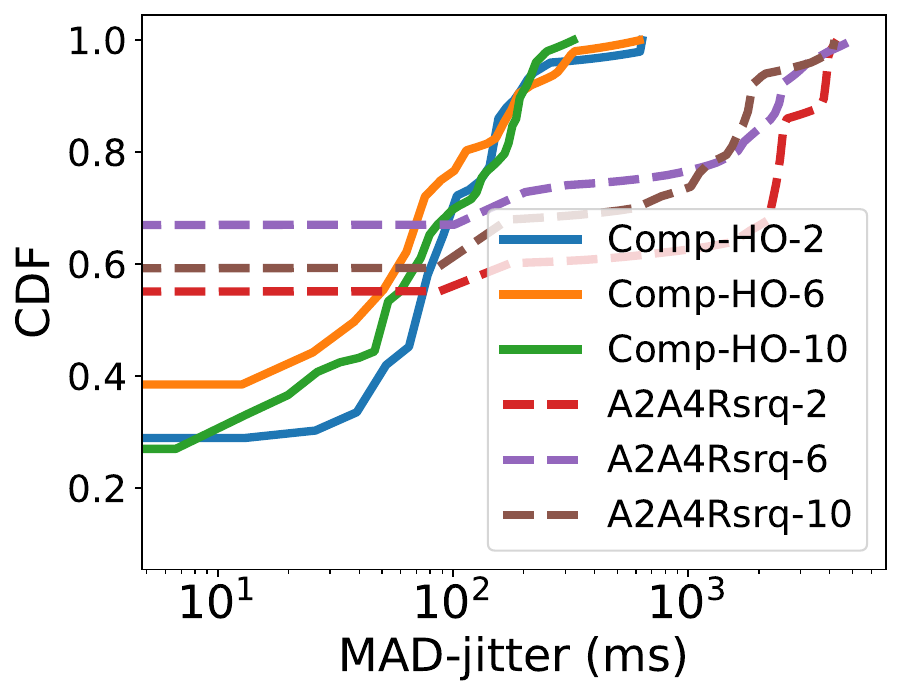}
      \caption{UE experienced MAD-jitter}
      \label{fig:jitterspeed}
    \end{subfigure}
    \caption{MEC processed packets, MEC-mobility packet loss ratio~(PLR), UE experienced delay and MAD-jitter (x-axes on log scale) for \textit{Comp-HO}, \textit{A2-A4-RSRQ}, \textit{A3-RSRP}, and \textit{NoHO} (no handoff) with different UE speeds (2, 6, and 10 m/s)}
    \label{fig:9}
\end{figure*}
\subsection{Transmission Performance}

\noindent\textbf{Improvement: }\autoref{fig:delay} and \autoref{fig:jitter} illustrate the cumulative distributions of UE experienced delay and jitter for all packets from UEs. We note that since the UE experienced delay distributions are very broad, we use a robust median-absolute deviation~(MAD) version of jitter. As shown, \textit{Comp-HO} improves the UE experienced delay with considerable improvements at the tail of the distribution. While, for jitter, \textit{Comp-HO} essentially smooths out the distribution, thus removing the large bifurcation seen in the other algorithms. \textit{Comp-HO} does this by better distributing the UEs across MECs and thus avoiding the dichotomy of ``lucky'' UE that happen to be in the areas with fewer competitors for channel and MEC resources. \autoref{fig:process} further supports this by showing that with \textit{Comp-HO} the MEC servers actually process more packets. Numerically, the mean UE experienced delay for \textit{Comp-HO} is 348\,ms compared to 1534\,ms, 1703\,ms, and 1667\,ms for \textit{A2-A4-RSRQ}, \textit{A3-RSRP}, and \textit{NoHO}, respectively.

\vskip 0.03in\noindent\textbf{Trade-off: }To illustrate the potential trade-off in using \textit{Comp-HO}, Figures 6c-7d illustrate the uplink and downlink SINR, transmission delay and jitter, therefore isolating transmission dynamics by removing processing dynamics. The results show that \textit{Comp-HO} has somewhat lower SINR and higher transmission delay and jitter, as would be expected given that \textit{Comp-HO} does not purely optimize SINR. However, the increases are relatively minor in comparison to the UE experienced delay and jitter decreases in the simulation. We also note that, as expected, the MEC-mobility packet loss of \textit{Comp-HO} is 4.3\% compared to 1.2\% for \textit{A2-A4-RSRQ}. This is the result of \textit{Comp-HO} more eagerly switching base stations given high delay MECs. However, much of this loss could be avoided in ETSI MEC networks that include MEC assisted user context transfer as the transfer could include data (packets) waiting for processing.

\begin{figure}
\begin{subfigure}{0.24\textwidth}
    \includegraphics[width=\linewidth]{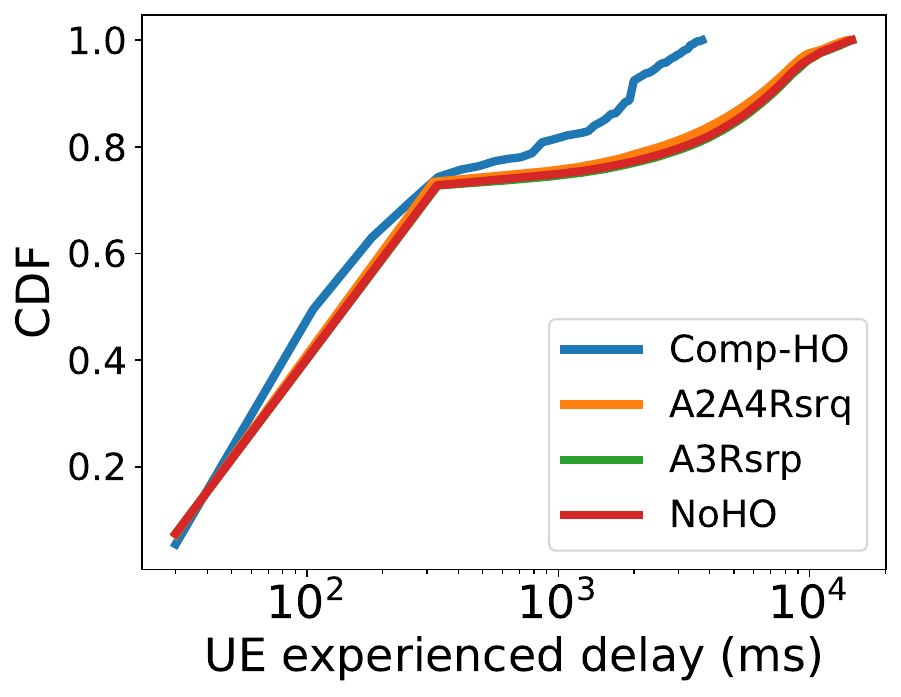}
      \caption{FPS: 30\,Hz}
      \label{fig:30ipt}
    \end{subfigure}
  \begin{subfigure}{0.24\textwidth}
    \includegraphics[width=\linewidth]{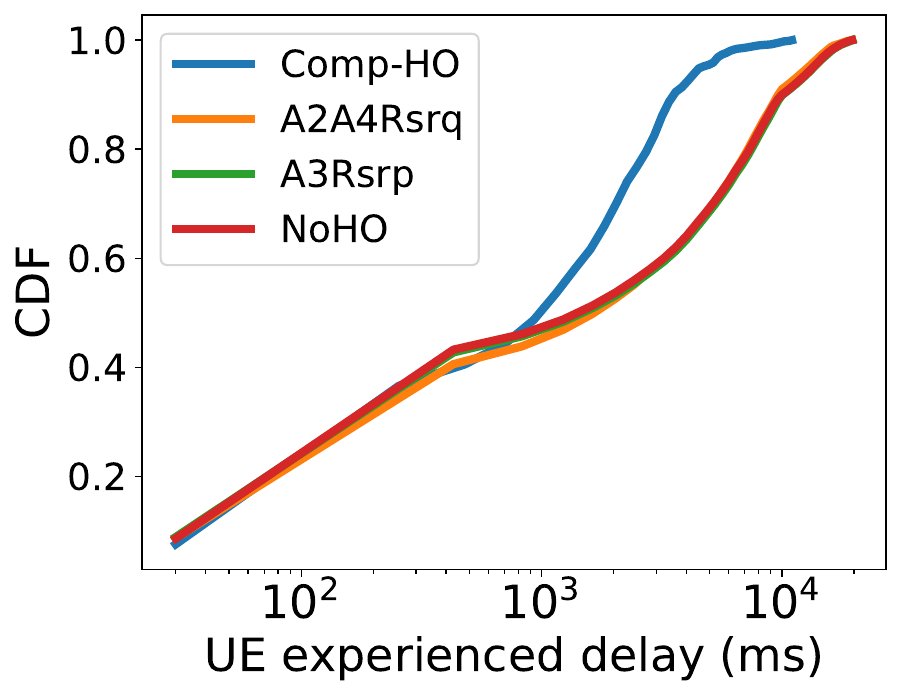}
      \caption{FPS: 50\,Hz}
      \label{fig:20ipt}
    \end{subfigure}
    \caption{UE experienced Delay with different FPS.}
    \label{fig:ipt}
\end{figure}
\subsection{Effects of Variations}
\noindent\textbf{Handoff rate:} For different effective handoff rates (as varied through changes in UE speed), \autoref{fig:plr} shows that \textit{Comp-HO} does suffer higher MEC-mobility packet loss ratio~(PLR) comparing with other algorithms with slower UEs. However with faster UEs, \textit{Comp-HO} maintains roughly the same PLR level while the other algorithms levels increase considerably. As such, \textit{Comp-HO} shows more robustness to higher UE speeds. Meanwhile, \autoref{fig:delayspeed} and \autoref{fig:jitterspeed} illustrate that higher handoff rates do improve the baseline algorithms\footnote{For brevity reasons the figures only show \textit{A2-A4-RSRQ}, however \textit{A3-RSRP} and \textit{NoHO} perform similarly.} performance relative to \textit{Comp-HO}. However, these improvements are only modest and the general patterns are similar. The improvements are actually partly an artifact of the increase in MEC-mobility packet loss since more long-delayed packets are lost and not included in the delay distributions.

\vskip 0.03in \noindent\textbf{FPS: }\autoref{fig:delay} together with \autoref{fig:30ipt} and \autoref{fig:20ipt} illustrate the delay distributions when using different frame rates (FPS), in other words different UE sending rates, within a common FPS range of MAR applications, i.e., 20\,Hz, 30\,Hz, and 50\,Hz, respectively. The results show that \textit{Comp-HO} improves user experienced delay in all FPS, though the degree of improvement varies likely due to an interplay of factors. 

\begin{figure}[!t]
\centering
\includegraphics[width=.6\linewidth]{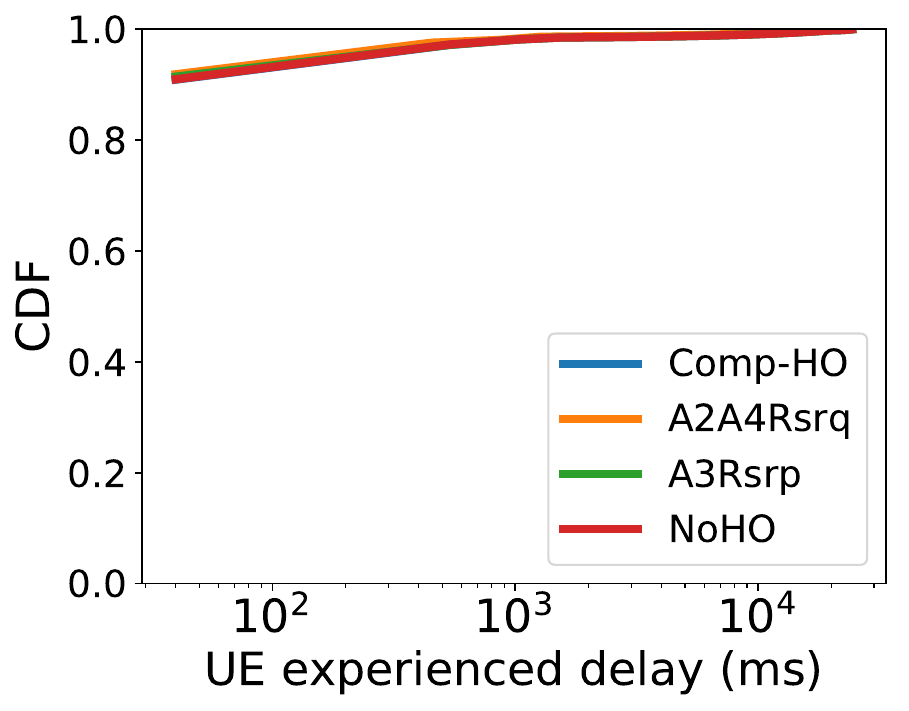}
\caption{UE experienced delay with Gauss-Markov mobility model.}
\label{fig:gauss}
\end{figure}

\vskip 0.03in \noindent\textbf{Mobility Models: } For the two different mobility models, Figures \ref{fig:delay} and \ref{fig:gauss} illustrate the UE experienced delay distributions of random waypoint and Gauss-Markov model respectively. As described before, with the baseline center-biased random waypoint model (with more heterogeneous user density and MEC loads) the \textit{Comp-HO} algorithm provides significant gains. However, with the Gauss-Markov model (with more homogeneous user density and MEC loads) the \textit{Comp-HO} algorithm provides very little benefit because there is less congestion at MEC servers and thus less need for load redistribution.

\begin{figure}[!t]
\centering
\includegraphics[width=.6\linewidth]{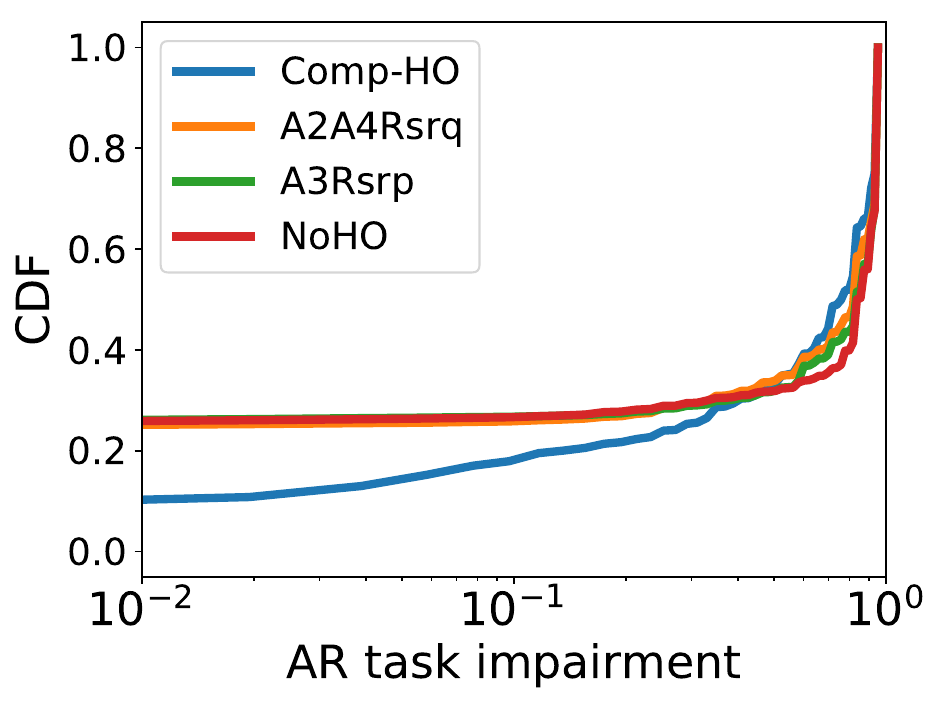}
\caption{AR task impairment distributions. An impairment score of one represents the maximum empirical performance while a score of zero represents the minimum, thus a higher score is better.}
\label{fig:impairment_dist}
\end{figure}
\subsection{Quality of Experience}
To illustrate the potential impact on user QoE, we utilise an existing AR task impairment model \cite{krogfoss2020} to transform the delay distribution values into AR task impairment scores. Specifically, an impairment score is a normalised score that quantifies the reduction in performance (from an empirical maximum) on a task or game (e.g., game score), specifically in this case, collaborate assembly of a virtual AR object like in Minecraft Earth (though also with physical object detection in our case). In other words, an impairment score of one represents the maximum empirical performance while a score of zero represents the minimum. \autoref{fig:impairment_dist} illustrates the impairment score distributions for the different handoff algorithms. We find that \textit{Comp-HO} significantly decreases the fraction of packets with full impairment (indicating the lowest performance level), thus suggesting that the delay improvements from \textit{Comp-HO} should translate into actual QoE gains during these types of AR tasks.

\begin{figure*}[!t]
  \begin{minipage}[b]{0.47\textwidth}
    \centering
    \includegraphics[width=\linewidth]{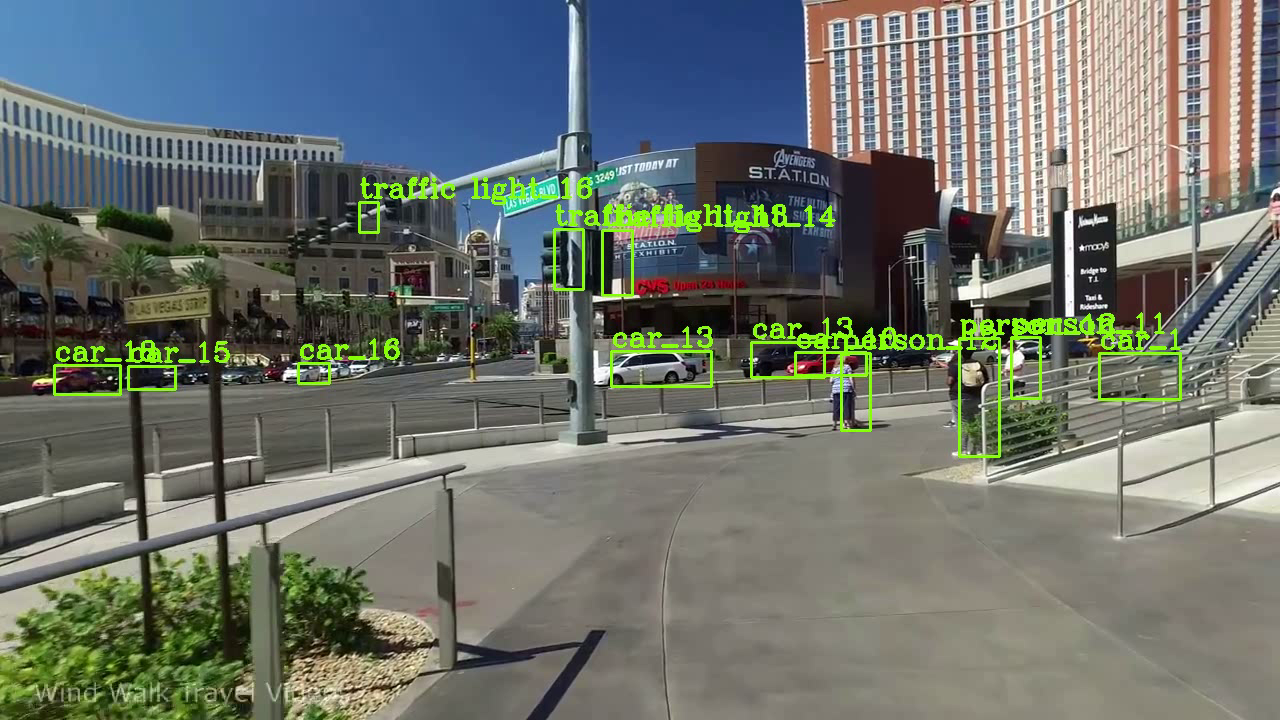}
      \caption*{Comp-HO}
      \label{Comp-HO}
  \end{minipage}
  \hfill
  \begin{minipage}[b]{0.47\textwidth}
    \centering
    \includegraphics[width=\linewidth]{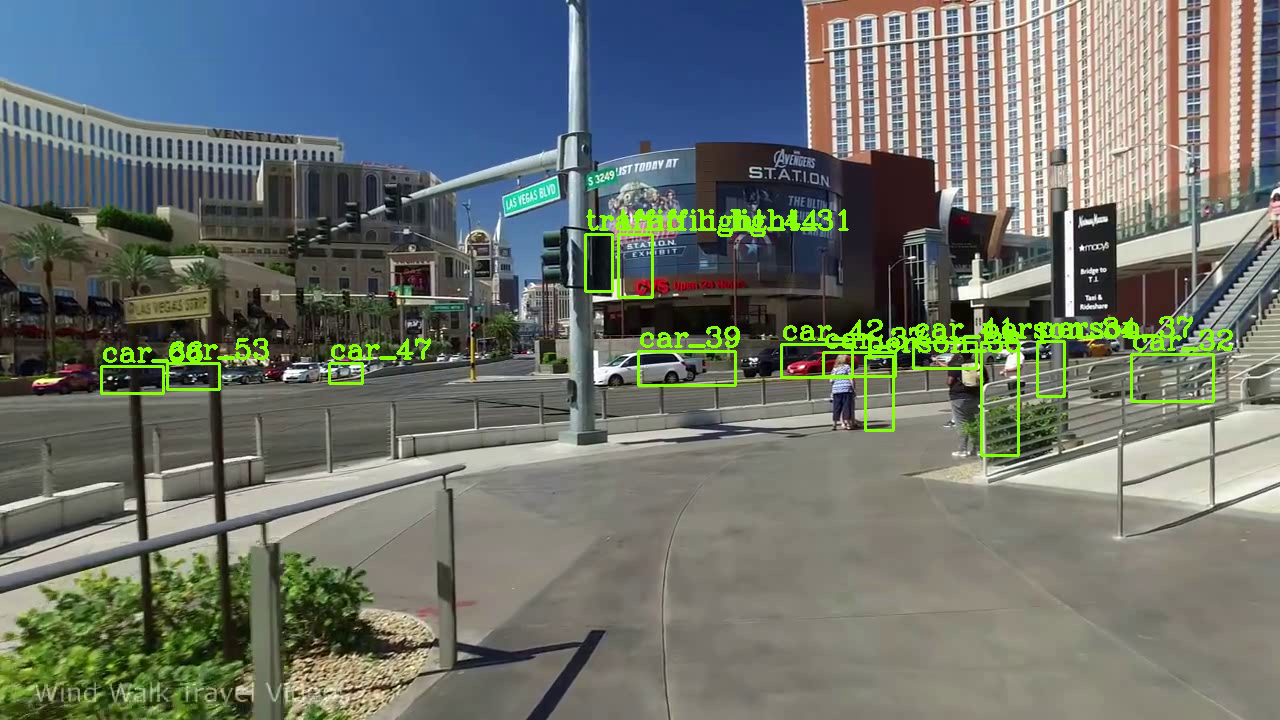}
      \caption*{A2A4Rsrq}
      \label{A2A4Rsrq}
  \end{minipage}
  
  \begin{minipage}[b]{0.47\textwidth}
    \centering
    \includegraphics[width=\linewidth]{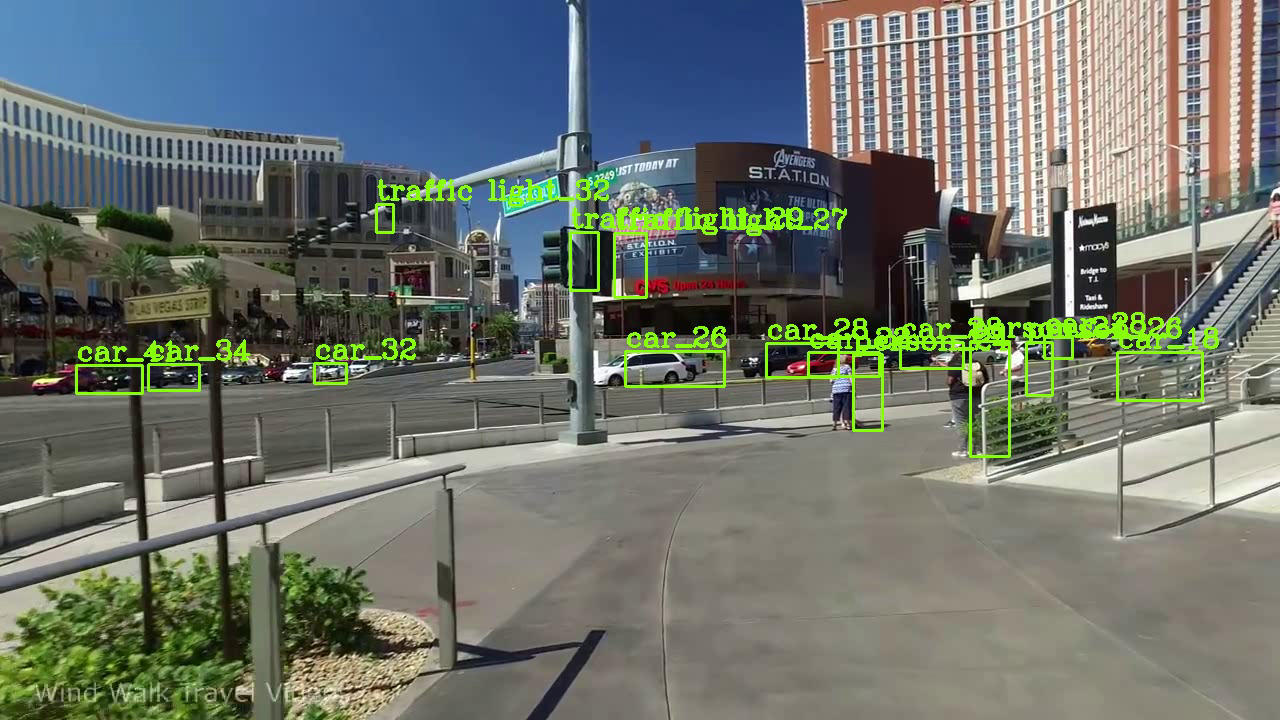}
      \caption*{A3Rsrp}
      \label{A3Rsrp}
  \end{minipage}
  \hfill
  \begin{minipage}[b]{0.47\textwidth}
    \centering
    \includegraphics[width=\linewidth]{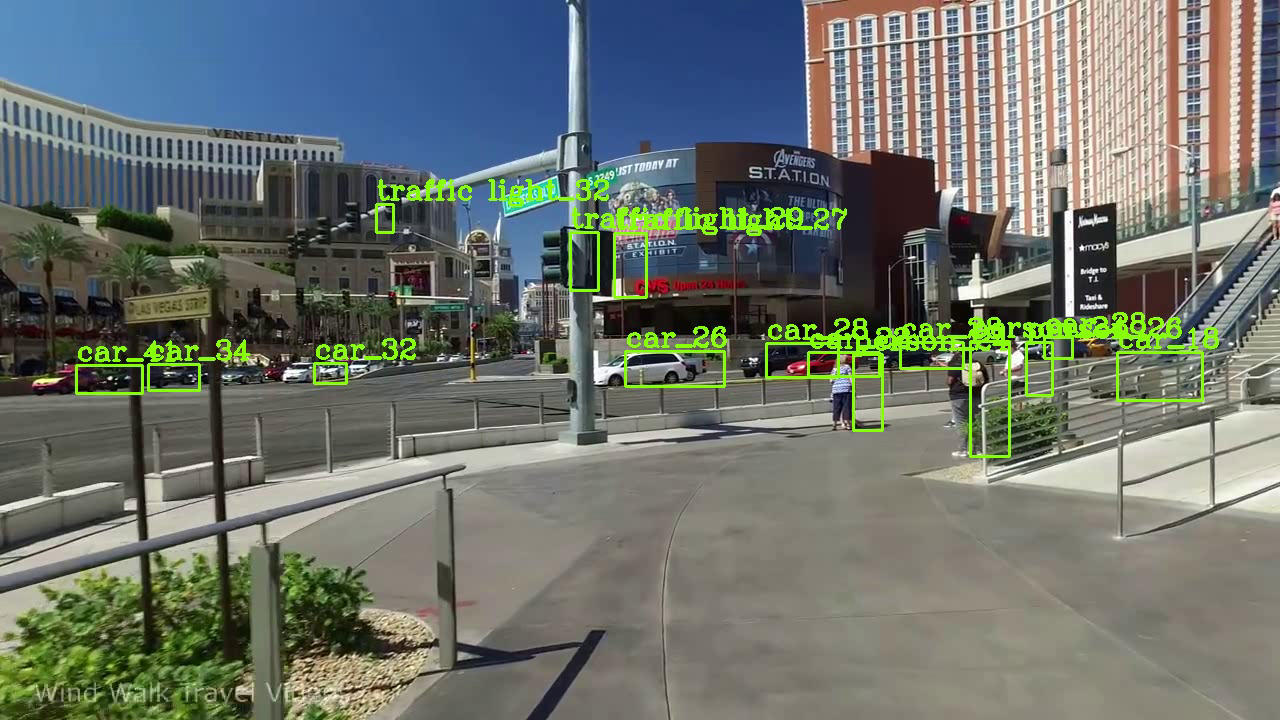}
      \caption*{NoOp}
      \label{NoOp}
  \end{minipage}
  
  \caption{YOLO\cite{yolov3} Demo for \textit{Comp-HO}, \textit{A2-A4-RSRQ}, \textit{A3-RSRP}, and \textit{NoHO}}
  \label{fig:demo}
\end{figure*}

\fu{
In \autoref{fig:demo}, we showcase the potential of \textit{Comp-HO} through an Object Detection Demo. To illustrate the impact of a handover algorithm on MEC applications, we simulate Object Detection App using YOLO\cite{yolov3} on a walk tour video\footnote{Video Clip from https://www.youtube.com/watch?v=uppuxvbW-w8}. While the top-left screenshot employs our proposed \textit{Comp-HO}, other screenshots use baseline algorithms respectively. For example, taking the car at the center, the bbox generated by \textit{Comp-HO} results in a shift of 13 pixels, whereas \textit{A2-A4-RSRQ}, \textit{A3-RSRP}, and \textit{NoHO} get a shift of 39, 26, 26 pixels respectively. Our demo demonstrates that \textit{Comp-HO} provides a more stable user experience by reducing delay for MEC applications in 5G scenarios.
}

\subsection{Takeaway} 
The simulation results show that \textit{Comp-HO} significantly improves the user experienced delay at the expense of a small increase in transmission delay (due to a decrease in signal strength). Additionally, \textit{Comp-HO} is more robust to different UE speeds and outperforms benchmark algorithms in different FPS.
Comparing with homogeneous user density and MEC loads distribution, \textit{Comp-HO} provides more improvement of user experienced delay in heterogeneous counterpart.

\section{Discussion}
\label{sec:limit}
In this work we proposed and evaluated \textit{Comp-HO} as a simple, effective, and standard-friendly computational handoff algorithm. In the future though, both \textit{Comp-HO} and our evaluation methods can still be improved in several aspects.

In terms of our evaluation, firstly, the 5G tesbed uses commercial BSs and thus due to technical and legal reasons we could not implement \textit{Comp-HO} directly into those BSs. Secondly and similarly, the custom MEC-enabled \texttt{NS-3} simulator could include some features like the consideration of user context migration between MEC servers (e.g., the migration of docker containers or VMs containing the current app state \cite{ma2017efficient,zavodovski2018icon}) and a more realistic MAR model at the application level.

In terms of improvements to \textit{Comp-HO} itself, the algorithm adds overheads such as collecting and sending the MEC load information. Thus, in cases with very few or very spatially homogeneous users, the algorithm will essentially act as a strongest SINR algorithm (since the MEC loads will be similar) but with an additional overhead and thus less efficient. Adding a threshold based on the number or distribution of UEs could remedy this by allowing base stations to fall back to traditional handoff algorithms during the mentioned conditions. 

Additionally, similar to having the \texttt{NS-3} simulator consider the user context migration in the evaluation, \textit{Comp-HO} could consider the user migration cost in the optimization directly (assuming differing migration costs between different MEC pairs).

In future work, we will integrate \textit{Comp-HO} and related algorithms into open Software Defined Radio (SDR) base stations, therefore allowing evaluations in a real physical deployment.
We also plan to further develop the MEC-enabled \texttt{NS-3} simulator to allow more comprehensive simulations. We will develop algorithms that consider different types of interference in 5G networks and the tradeoff between the handoff rates and the level of interference in the network. 
In addition, we will develop a predictive capability algorithm allowing us to predict better mobility so resources can be pre-allocated before a handoff event even occurs. 

\section{Conclusion}
\label{sec:conclusion}
The combination of 5G and MEC allows novel services and improved experiences in areas like MAR and virtual reality among others. Towards this goal and with a focus on mobility, this work studied the issue of \textit{MEC HO} in 5G and proposed a handoff algorithm, \textit{Comp-HO}, that considers both network signal strength and nearby MEC server load. We conducted the 5G MEC testbed measurements with a MAR prototype and utilized the collected results to set up large scale simulations with a custom MEC-enabled \texttt{NS-3} simulator. The simulation results illustrated that \textit{Comp-HO} improves the end-to-end delay compared to benchmark handoff algorithms in MEC scenarios. As far as we know, this is one of the first efforts to optimize \textit{MEC HO} for 5G.

\bibliographystyle{IEEEtran}
\bibliography{IEEEabrv,MM/references}

\end{document}